# STREAM: An Objective-Driven and Uncertainty-Aware Framework for Industrial Energy Data Acquisition

Zhipeng Ma[1[0000-0002-4049-539X]], Bo Nørregaard Jørgensen[1[0000-0001-5678-6602]], Zheng Grace Ma[1[0000-0002-9134-1032]]

[1] SDU Center for Energy Informatics, the Maersk Mc-Kinney Moller Institute, Faculty of Engineering, University of Southern Denmark, DK-5230 Odense, Denmark

{zhma, bnj, zma}@mmmi.sdu.dk

**Abstract.** Industrial energy management requires datasets that connect energy use with equipment states, production batches, material flows, and process conditions. However, conventional acquisition workflows commonly emphasize connectivity and storage without verifying whether accessible signals satisfy the requirements of a defined energy-performance assessment. This paper presents STREAM, an objective-driven and uncertainty-aware framework comprising Specification of Objectives, Technical Requirements, Resource Mapping, Extraction from Sources, Archival Metadata, and Migration to Database. STREAM is the central workflow: objective-to-data traceability is its end-to-end output, while measurement, temporal, contextual, and processing uncertainty are assessed across all six stages. Compared with the original conceptual STREAM sequence, this paper adds stage-level artifacts, minimum-evidence gates, source-suitability rules, a metadata template, an uncertainty rubric, and case-specific traceability matrices. The framework is validated through two industrial batch-process cases: induction-furnace melting in a foundry and cheese-powder drying using SCADA and production-order data. The results demonstrate that data accessibility is not equivalent to analytical suitability and show how STREAM supports transparent decisions about immediate data use, analytical restrictions, and prioritized infrastructure improvements.

## 1 Introduction

Industrial decarbonization requires production systems to use energy efficiently and to explain where, when, and under which operating conditions energy is consumed [1-2]. In energy-intensive processes, performance depends on equipment states, production volumes, material properties, schedules, temperatures, and control decisions. Aggregate electricity or thermal-energy measurements are therefore insufficient without a reliable process context [2-3].

Industrial digitalization has increased the availability of data from sensors, energy meters, SCADA systems, industrial Internet of Things platforms, historians, and pro-duction databases [4-5]. These data can support monitoring, benchmarking, anomaly detection, flexibility assessment, and decision support [4-6]. However, data availability does not ensure analytical suitability. Accessible signals may still lack the variables, timing, context, calibration evidence, or transformation history required for a specific assessment [6-8]. Factory data infrastructures are typically designed for control, monitoring, maintenance, or safety rather than energy-performance assessment [4-5]. As a result, energy indicators may be affected by inconsistent timestamps, heterogeneous sampling rates, unresolved sensor-to-asset relationships, unclear system boundaries, missing batch links, and incomplete metadata [5, 7, 9-11].

The scientific gap addressed in this paper is the lack of an operational acquisition framework for verifying whether industrial sensors and production data are fit for a defined energy objective before indicators are calculated or decisions are made. Unreliable data can produce incorrect benchmarks, misleading best-practice identification, and weak confidence in data-driven decisions [12-13].

This paper therefore develops and evaluates a framework that links energy objectives to required variables, physical and digital sources, metadata, uncertainty records, and analytical repositories. The research question is: How can industrial energy-data acquisition be structured so that objective definition, variable specification, source verification, uncertainty documentation, and database migration form a traceable chain?

To answer this question, STREAM is employed as the organizing framework. Data extraction is embedded in the Extraction from Sources and Migration to Database stages, objective-to-data traceability is the output of the six-stage workflow, and uncertainty assessment supports source-suitability decisions across all stages. The initial STREAM concept in [14] introduced the six-stage sequence at a conceptual level. This paper operationalizes it through stage-specific artifacts, minimum-evidence gates, failure actions, variable-suitability classes, a mandatory metadata structure, a four-category uncertainty rubric, and two industrial applications with variable-source and uncertainty matrices. The contribution is therefore an operational requirements-to-evidence procedure for industrial energy-data acquisition.

The framework is evaluated through induction-furnace melting in a foundry and cheese-powder drying in a dairy process. The purpose of the evaluation is not to claim energy savings, but to test whether STREAM can diagnose data readiness and prepare traceable datasets for monitoring, benchmarking, and decision support.

The remainder of the paper is organized as follows. Section 2 reviews related work. Section 3 describes the STREAM framework. Section 4 presents the two industrial case studies. Section 0 reports the cross-case results. Section 6 discusses the findings, contributions, limitations, and future work, and Section 7 concludes the paper.

# 2 Background and Related Work

## 2.1 Industrial Sensor-Data Acquisition and Management

Industrial energy and process data are generated by sensors, energy meters, programmable logic controllers, SCADA systems, industrial IoT platforms, and production da-tabases [4-5]. Because these data are distributed across heterogeneous systems, their integration is often affected by inconsistent timestamps, formats, provenance, and governance practices [4-5, 15].

Most acquisition architectures emphasize connectivity, communication, transmission, and storage. Although necessary, these capabilities do not ensure suitability for a specific energy-management objective. An accessible signal may still have inadequate sampling, missing calibration evidence, unclear asset relationships, or insufficient process context [7, 11]. Data accessibility should therefore be distinguished from analytical suitability [8].

Metadata provides the context required to interpret and reuse observations, including source and asset identifiers, units, sampling intervals, measurement characteristics, calibration status, installation context, and transformation history [10, 16]. Standards such as the Semantic Sensor Network ontology support relationships among sensors, observations, measured properties, platforms, and system characteristics [6, 10, 16]. However, industrial metadata are often fragmented across documentation, engineering systems, SCADA configurations, databases, and expert knowledge [5, 9]. Integrating metadata into the acquisition workflow is therefore essential for traceability, interoperability, and reliable reuse in analytical repositories [9, 16].

### 2.2 Energy-Performance Indicators and Objective-Driven Requirements

Industrial energy data are only analytically meaningful when they are connected to clearly defined energy-performance indicators and decision uses. Examples include batch electricity consumption, specific energy consumption, energy use per unit of product, thermal energy per unit of water removed, and deviations from an energy baseline [3]. Each indicator imposes different requirements on system boundaries, time resolution, normalization variables, process-state information, and acceptable uncertainty.

For example, batch-level specific energy consumption requires energy values, material or product quantity, batch start and end times, equipment identifiers, and process-state information. A signal that is sufficient for dashboard monitoring may therefore be insufficient for batch benchmarking or optimization if it cannot be aligned with the production context. Objective-driven acquisition makes these requirements explicit before the available sensor tags are evaluated.

This objective-driven view provides the first link in the STREAM logic. It separates the question of data accessibility from the stronger question of analytical suitability and provides the basis for classifying sources as suitable, conditional, unsuitable, or unavailable for a specified indicator.

### 2.3 Data Uncertainty

Measurement uncertainty, in the metrological sense, describes the dispersion of values that could reasonably be attributed to a measured quantity and should be considered together with calibration, accuracy, resolution, drift, installation conditions, and environmental effects [17-18]. In industrial sensor systems, uncertainty is not limited to measurement accuracy. Temporal uncertainty arises from irregular sampling, clock dif-ferences, communication delays, and uncertain batch boundaries. Contextual uncertainty arises when observations cannot be reliably linked to the corresponding equipment, production order, material, recipe, or operating state. Processing uncertainty is introduced when raw observations are filtered, resampled, interpolated, aggregated, converted, or otherwise transformed [7, 9, 11].

These uncertainty types influence energy analyses differently. General monitoring may tolerate moderate timestamp or contextual uncertainty, whereas batch-level benchmarking requires reliable synchronization of energy, production, and process data [12-13]. Similarly, interpolation may be acceptable for visualization but inappropriate for precise energy-performance assessment [6, 11].

Data quality and uncertainty are related but distinct. Data-quality indicators describe observable dataset properties, such as missing records, duplicates, and sampling irregularities [6, 19]. Uncertainty concerns incomplete knowledge about the value, timing, context, or transformation of an observation [9]. A complete dataset may still be uncertain when calibration evidence, batch linkage, or transformation history is unavailable. These distinctions motivate an acquisition framework that documents not only whether data exist, but also whether there is sufficient evidence to use them for the intended energy-performance assessment.

# 3 The Objective-Driven and Uncertainty-Aware STREAM Framework

## 3.1 Overview

Building on the STREAM concept illustrated in [14], this paper operationalizes STREAM as an objective-driven and uncertainty-aware framework for industrial energy data acquisition. As shown in Fig. 1, it comprises six stages: Specification of Objectives (S), Technical Requirements (T), Resource Mapping (R), Extraction from Sensors (E), Archival Metadata (A), and Migration to Database (M) [14]. These stages form a traceable chain from the energy-performance question to the repository used for analysis.

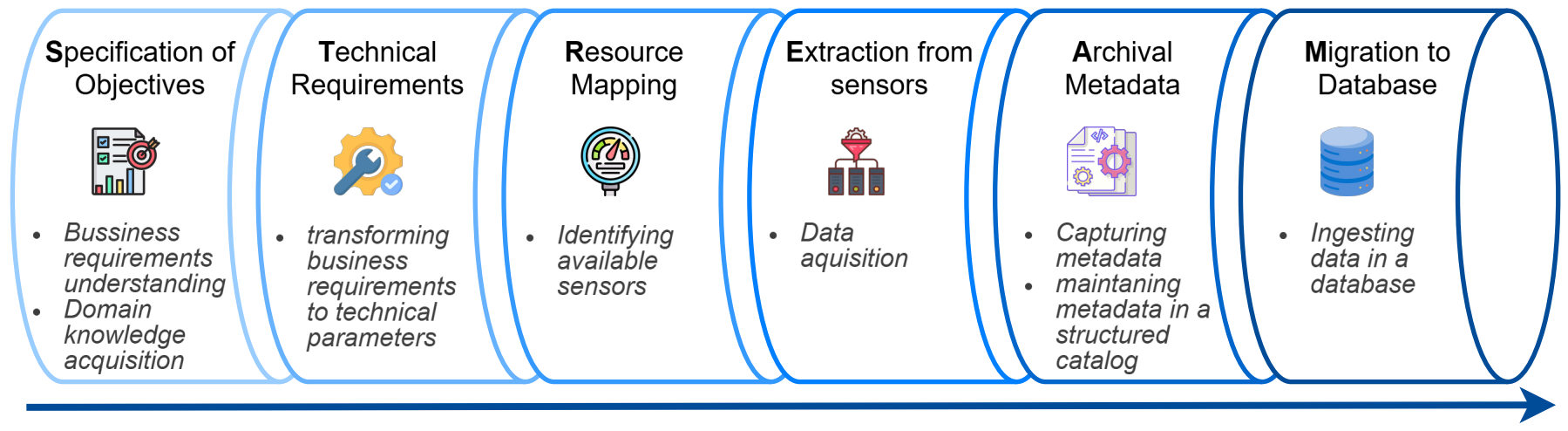

**Fig. 1.** STREAM acquisition workflow linking objectives, requirements, physical sources, metadata, and repositories; uncertainty management cuts across all stages [14].

STREAM starts from the decision or energy-management purpose rather than from the signals already available. It links objectives with physical measurements, contextual information, uncertainty records, and analytical infrastructure, thereby evaluating availability, traceability, source suitability, and reliability for the specified task.

Uncertainty is addressed during all six stages because problems discovered after database migration are often difficult to correct. STREAM therefore records uncertainty evidence while objectives are defined, sources are selected, observations are extracted, metadata are archived, and records are migrated.

Table 1 summarizes the operational protocol. Its decision logic is conservative: a source is not considered analysis-ready until the required evidence is verified or its remaining limitations are documented.

**Table 1.** Operational structure of STREAM.

| Stage | Guiding question | Artifact | Minimum evidence | Failure action |
|---|---|---|---|---|
| S | What decision or indicator is supported? | Objective specification | Indicator, boundary, use, owner | Refine or split the objective |
| T | What variables and tolerances are needed? | Measurement specification | Variable, unit, range, sampling, accuracy, alignment and missingness tolerance | Add, relax, or prioritize requirements |
| R | Which source provides each variable? | Resource map | Source ID, asset link, access, calibration and status | Add sensor, use proxy, combine sources, or record gap |
| E | Can data be extracted traceably? | Extraction log and raw dataset | Source timestamp, extraction timestamp, source ID, unit, quality flag | Repair, flag failures, preserve raw data |
| A | What metadata makes data interpretable? | Metadata catalog | Required metadata complete or marked unknown | Complete metadata, assign IDs, document assumptions |
| M | Can data be reused reliably? | Validated repository | Schema, units, duplicate checks, normalized timestamps, uncertainty flags | Reject, reprocess, or quarantine records |

STREAM is applied sequentially but iteratively. The practitioner completes the objective specification for one indicator, translates it into critical and non-critical requirements, and maps each critical variable to one or more candidate sources. Only suitable or explicitly conditional sources proceed to extraction. Extraction failures and processing steps are logged, metadata are checked against the mandatory fields, and the resulting observations and evidence are migrated together. A failed evidence gate triggers the failure action in Table 1 and returns the process to the earliest affected stage. The final deliverable is therefore a dataset accompanied by a traceability record explaining why each variable is suitable, conditional, unsuitable, or unavailable.

### 3.2 The Six STREAM Stages

**Specification of Objectives.** This stage defines the energy or operational decision to be supported, the process boundary, the intended indicator, and the analytical use [19].

Typical objectives include reducing specific energy consumption, comparing production batches, detecting abnormal demand, or assessing flexibility. The output is an objective specification that links the decision purpose to a measurable indicator with explicit physical and temporal boundaries.

**Technical Requirements.** This stage translates each indicator into required variables, units, expected ranges, sampling intervals, accuracy, resolution, missing-data tolerance, and temporal-alignment tolerance [19].

Critical requirements are separated from non-critical requirements: the former determine feasibility, whereas the latter affect confidence or convenience.

The output is a measurement specification used to evaluate existing sources.

**Resource Mapping.** This stage identifies and documents candidate sensors, control-system tags, databases, production systems, and complementary records, and maps each source to the technical requirement it may satisfy.

For every candidate source, the practitioner records the source identifier, measured quantity, asset relationship, installation context, range, accuracy, resolution, sampling behavior, calibration status, accessibility, and operational status. The evidence is then compared with the critical and non-critical requirements defined in the T stage.

The outputs include a source-to-requirement map, a measurement gap list, and a suitability classification for every required variable.

A source is suitable when it satisfies all critical requirements and provides the mandatory metadata needed for interpretation. It is conditional when its limitations can be managed through a documented mitigation or analytical restriction. It is unsuitable when at least one critical requirement fails and no acceptable mitigation exists for the intended indicator, and unavailable when no source exists within the defined boundary.

**Extraction from Sensors.** This stage retrieves observations from approved sources. Acquisition mode and technical implementation are documented separately.

Acquisition modes include continuous ingestion, scheduled polling, event-triggered retrieval, direct database extraction, and file-based batch transfer. Technical implementations may include OPC UA or MQTT gateways, REST interfaces, SQL queries, historian connectors, or CSV exports. For each source, the extraction log records the mode, implementation, frequency, source and extraction timestamps, identifiers, units, clock source, quality flags, error handling, and processing status.

Communication failures, missing intervals, duplicated records, irregular sampling, and delayed transmission are recorded rather than concealed. Raw observations are preserved whenever possible, while corrected, resampled, interpolated, aggregated, or estimated values receive an explicit processing status.

The output is a timestamped raw dataset and extraction log with traceable source, timing, quality, and processing information.

**Archival Metadata.** This stage augments extracted observations with the contextual, quality, provenance, and uncertainty metadata required for interpretation and reuse.

The metadata catalog is linked to observations through persistent source, asset, batch, order, and dataset identifiers. A detailed template is provided in Appendix Table A1. Mandatory fields are defined before application and versioned with the dataset. Fields that cannot be verified are recorded as "unknown", together with the responsible actor and planned mitigation, rather than being completed through undocumented assumptions.

**Migration to Database.** This stage transfers observations, metadata, quality records, and uncertainty information into a persistent repository, such as a relational, time-series, document-based, or industrial data-platform database.

Migration includes schema validation, identifier verification, timestamp normalization, unit checking, duplicate detection, error logging, and ingestion monitoring, while keeping measured observations distinguishable from corrected, aggregated, interpolated, or estimated values.

The resulting repository supports queries by time, source, asset, measured quantity, batch, process state, quality status, and processing status.

### 3.3 Cross-Cutting Uncertainty Management

Uncertainty management is implemented as an evidence-based assessment rather than as a general checklist. For every required variable, the practitioner records the evidence inspected, the relevant uncertainty category, a low, moderate, or high/unknown risk level, the reason for the rating, its effect on the intended indicator, and the required mitigation or analytical restriction. Appendix Table A2 provides the assessment rubric.

Measurement uncertainty concerns sensor accuracy, resolution, calibration, drift, installation conditions, and environmental effects; STREAM manages it through measurement requirements, sensor evaluation, and calibration metadata.

Temporal uncertainty concerns clock drift, communication delays, irregular sampling, missing intervals, and uncertain boundaries; it is managed through synchronization tolerances and timestamp evidence.

Contextual uncertainty concerns weak links between observations and equipment, batch, recipe, material, or operating state; it is managed through source-to-asset mapping and process metadata.

Processing uncertainty concerns filtering, resampling, interpolation, aggregation, unit conversion, and other transformations; STREAM requires raw-data preservation and transformation records.

The categories are interdependent: missing observations create temporal uncertainty, interpolation creates processing uncertainty, and an accurate value can remain analytically uncertain when its asset, batch, or operating state is unknown. Quantitative esti-mates are used when calibration certificates, sensor specifications, synchronization tests, or validation data are available; otherwise, the evidence gap is recorded explicitly through the qualitative rubric.

The suitability decision follows a conservative rule. A variable is suitable when all critical requirements are satisfied, and no relevant uncertainty category has unresolved high or unknown risk. It is conditional when a documented mitigation or restriction makes the intended use defensible. It is unsuitable when a critical requirement fails without an acceptable mitigation, and unavailable when no source exists. Indicator readiness is determined by the least favorable classification among its critical variables.

For batch-specific energy consumption, for example, the critical variables include energy, material mass, and batch boundaries. Missing meter calibration, uncertain mass evidence, or unreliable boundaries can therefore restrict the complete indicator even when the electricity signal itself is accessible. Data with moderate uncertainty may still support trend monitoring, whereas benchmarking, optimization, automated control, or formal reporting generally requires stricter evidence.

# 4 Case Studies

Two industrial batch-process cases are examined using the same operational proto-col and suitability rule. They test STREAM in different data environments rather than quantify energy savings: the first relies mainly on IoT-based furnace data, whereas the second integrates SCADA and production-order records.

## 4.1 Case Study 1

The first case examines induction-furnace melting in a metal foundry, one of the most energy-intensive foundry operations [20-21]. The boundary extends from material charging to completion of molten-metal transfer, including heating, possible composition adjustment, and ladle transfer.

The existing IoT infrastructure records process and energy data. The objective is to relate electricity consumption to production and process conditions using indicators such as batch electricity use, specific electricity consumption, melt duration, and temperature trajectory. The case tests whether accessible IoT data can be connected to material, furnace-state, and batch context before benchmarking.

**Table 2.** Overview and data-readiness focus of Case Study 1.

| Element | Description |
|---|---|
| Industrial process | Induction-furnace melting in a metal foundry |
| Production mode | Batch production with continuously recorded sensor data |
| Main equipment | Electrically powered induction furnaces and ladle transfer |
| Energy carriers | Electricity |
| Existing systems | IoT infrastructure connected to furnace sensors and control-system records |
| Data sources assessed | Electricity meter, furnace-temperature sensor, material or charge records, equipment-state records, and batch records |
| Candidate indicators | Batch electricity use, specific electricity consumption, melt duration, temperature-to-target behavior, and status energy |
| Critical readiness requirements | Calibration evidence, timestamp alignment, furnace-to-batch linkage, material quantity, and processing-state definitions |
| Intended application | Energy benchmarking, monitoring, and decision support |

**STREAM application.** The S stage defined batch-level electricity and process-performance indicators. T translated them into five critical variable groups: electricity consumption, furnace temperature, material quantity, batch boundaries, and equipment status. R mapped these groups to IoT meters, sensors, control-state tags, and production records. E extracted timestamped observations while preserving source identifiers and raw state changes. A augmented the records with furnace, batch, material, calibration, boundary, and transformation metadata. M validated identifiers, units, timestamps, duplicates, and processing status before loading the analytical repository.

### 4.2 Case Study 2

The second case examines cheese-powder production, focusing on the energy-intensive drying tower. Prepared feed is atomized and exposed to heated air; the resulting powder is separated, collected, and transferred for further handling. Although production is batch-based, sensor data are recorded continuously.

Electricity powers pumps, fans, atomization, and auxiliary equipment, while thermal energy heats the drying air. The objective is to relate electricity and heating consumption to production conditions using indicators such as electricity per order, thermal energy per unit of product, and recipe- or batch-normalized drying performance. The case tests synchronization and contextual linkage between continuous SCADA signals and discrete production-order records.

**Table 3.** Overview and data-readiness focus of Case Study 2.

| Element | Description |
|---|---|
| Industrial process | Cheese-powder production with focus on the drying tower |
| Production mode | Batch production with continuously recorded sensor data |
| Main equipment | Drying tower, atomization equipment, fans, pumps, and auxiliary systems |
| Energy carriers | Electricity and heating |
| Existing systems | SCADA system and production database |
| Data sources assessed | SCADA energy and process tags, heating records, humidity, temperature and pressure sensors, feed or product quantity records, and order database |
| Candidate indicators | Electricity per order, heating per order, energy per kg powder, where available, and recipe-normalized energy use |
| Critical readiness requirements | Heating data quality, SCADA timestamp synchronization, order and recipe linkage, product quantity, and transformation history |
| Intended application | Energy benchmarking, monitoring, and decision support |

**STREAM application.** The S stage defined electricity- and heating-related indicators at the order, product, and recipe level. T specified six variable groups: electricity demand, heating demand, process conditions, feed or product quantity, batch timing, and equipment status with transformation records. R mapped these groups to SCADA tags, heating records, and the production database. E extracted SCADA and order data separately while retaining their source timestamps and clock information. A augmented the data with tag, order, recipe, quantity, time-zone, and heating-derivation metadata. M harmonized the two schemas, checked units and timing, and preserved raw and derived values in the repository.

# 5 Results

This section reports and evaluates the two STREAM applications as data-readiness assessments rather than energy-performance claims. The evaluation uses source-mapping coverage, suitability distribution, uncertainty-category coverage, and corrective-action coverage derived from the completed artifacts in Appendix Tables A3-A6.

## 5.1 Results of Case Study 1

**Objective-to-Data Traceability.** STREAM maps all five required variable groups in Case Study 1 to candidate physical or digital sources. Electricity consumption, furnace temperature, batch timing, and equipment status are conditionally suitable, while material quantity is conditional with restricted analytical use. Thus, none of the five groups is fully suitable for unrestricted batch-level benchmarking. Detailed evidence is provided in Appendix Table A3.

The mapping confirms that extraction, metadata augmentation, and migration are technically feasible, but calibration, timestamp alignment, material evidence, batch linkage, and state definitions remain critical evidence gaps. A corrective action was specified for each of the five variable groups.

Fig. 2 illustrates examples of collected sensor data.

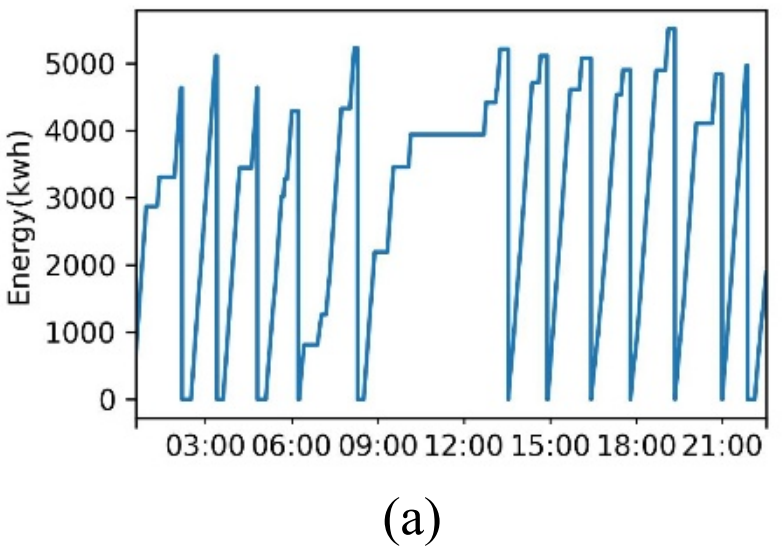

(a)

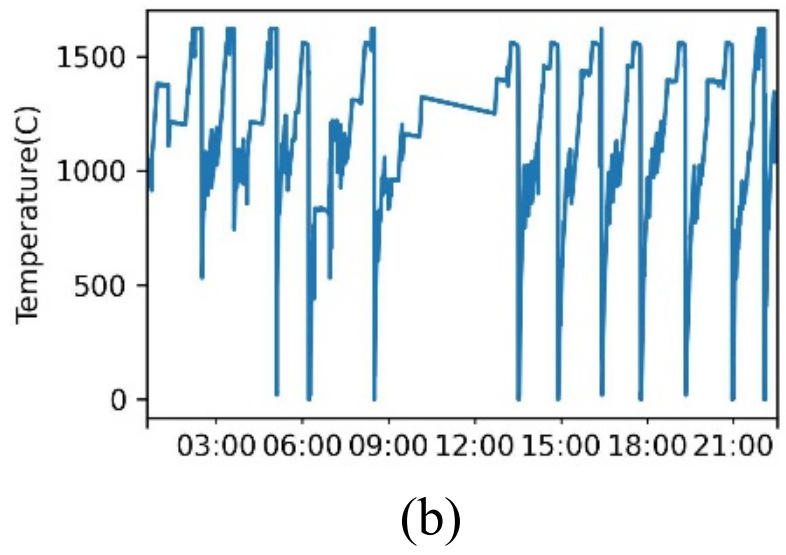

(b)

**Fig. 2.** Representative induction-furnace signals from the IoT system: (a) electricity signal and (b) furnace-temperature signal.

**Data-Uncertainty Assessment.** Gaps are identified in all four uncertainty categories: measurement, temporal, contextual, and processing. The detailed matrix is provided in Appendix Table A4.

The foundry dataset is therefore suitable for trend monitoring and exploratory analysis, but batch benchmarking remains conditional. STREAM changes the final analytical treatment by requiring uncertain batches to be restricted or excluded rather than assigning energy through undocumented boundary or material assumptions.

### 5.2 Results of Case Study 2

**Objective-to-Data Traceability.** STREAM maps all six required variable groups in Case Study 2 to candidate sources. Electricity demand, process conditions, feed or product quantity, batch timing, and equipment status are conditionally suitable, while heating demand is conditional with restricted analytical use. None of the six groups is fully suitable for unrestricted order- or recipe-level benchmarking. Detailed evidence is provided in Appendix Table A5.

The assessment confirms that SCADA and production-order data can be extracted and migrated, but heating-data quality, clock synchronization, order and recipe linkage, quantity evidence, and transformation provenance remain limiting. A corrective action was specified for each of the six variable groups.

Fig. 3 shows examples of collected sensor data.

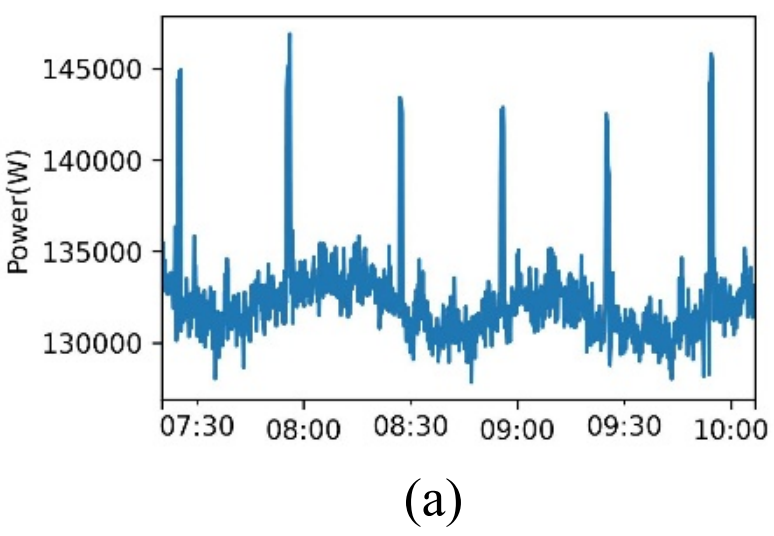

(a)

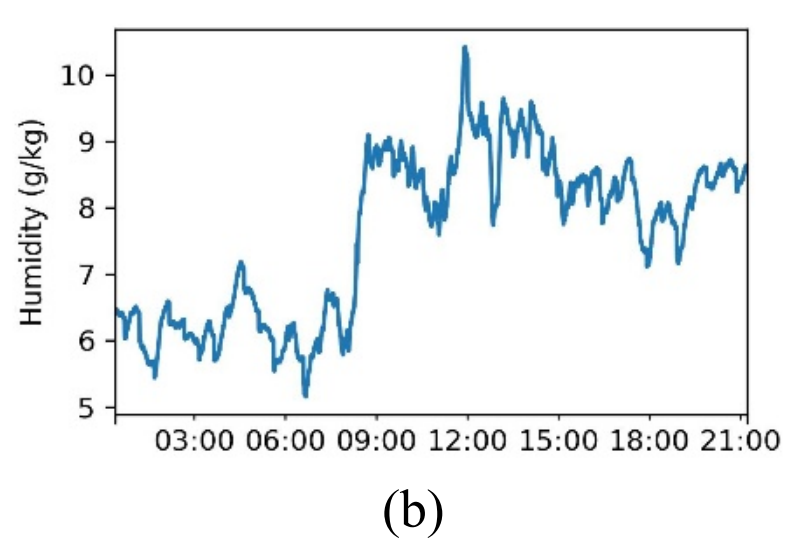

(b)

**Fig. 3.** Representative drying-tower signals from the SCADA infrastructure: (a) power signal and (b) humidity signal.

**Data-Uncertainty Assessment.** Gaps are again identified in all four uncertainty categories. The detailed matrix is provided in Appendix Table A6.

In the drying case, incomplete calibration evidence, low-quality heating data, sampling-rate differences, partial order, recipe, or quantity links, and transformation history reduce readiness for batch- or recipe-level benchmarking. STREAM shows that analytical readiness depends on system integration, contextual linkage, and transformation provenance rather than extraction alone.

## 5.3 Cross-Case Evaluation

Across both cases, all six STREAM stages produced traceable links from energy objectives to analytical repositories. Table 4 summarizes the recurring result: STREAM is feasible in both IoT- and SCADA-based environments, but it should be understood as a diagnostic framework that makes the evidence required for trustworthy indicators visible.

**Table 4.** Cross-case data-readiness summary.

| Finding | Case study 1 | Case study 2 | Implication |
|---|---|---|---|
| Energy data availability | IoT-based electricity and process signals are available. | SCADA-based electricity, heating, and process data are available. | Energy-data extraction is feasible in both data architectures. |
| Production context | Batch, material, and furnace-state linkages are incomplete. | Order, recipe, quantity, and operating-state linkages are incomplete. | Reliable benchmarking requires stronger integration of energy and production contexts. |
| Calibration and validation evidence | Evidence for key electricity and temperature sensors is limited. | Calibration evidence and heating-data quality records are limited. | Measurement reliability must be documented before using derived indicators. |
| Temporal alignment | Sampling rates and batch-boundary definitions varied across data sources. | SCADA records and order timestamps required synchronization. | Batch-level indicators require explicit time-alignment and boundary rules. |
| Transformation history | Resampling, aggregation, and unit conversion are not fully documented. | Resampling, derived heating values, and aggregation are not fully documented. | Raw data and transformation records should be preserved for traceability. |

The cross-case results clarify what STREAM adds to an industrial energy assessment. First, the same accessible signal may support one decision but not another: energy and temperature traces are adequate for monitoring trends, whereas batch benchmarking also requires reliable material quantities, batch boundaries, and equipment-state records. Second, uncertainty is not confined to sensor accuracy. In both cases, the main risks emerged at the interfaces between systems, where meters, process tags, order records, and transformations do not always share a complete time base or contextual link. STREAM therefore turns acquisition from a data-transfer activity into a readiness assessment. Its output is a justified classification of whether each required variable is suitable, conditional, unsuitable, or unavailable for the defined assessment. This prevents incomplete datasets from being used as un-questioned evidence in dashboards, benchmarks, or decision-support tools.

# 6 Discussion

The discussion interprets the data-readiness findings, compares STREAM with related approaches, states contributions, and summarizes implementation implications, validity threats, and limitations.

## 6.1 Comparison with Existing Approaches

The cases show that the central acquisition challenge is not connectivity alone but verifying source suitability for the intended assessment. Table 5 compares STREAM with pipeline, metadata, provenance, and energy-data protocol approaches using criteria aligned with this purpose. The comparison concerns procedural coverage rather than algorithmic performance.

**Table 5.** Comparison of data-collection and acquisition approaches.

| Approach | Objective alignment | Variable traceability | Metadata/provenance | Uncertainty treatment | Industrial-energy validation |
|---|---|---|---|---|---|
| Industrial data pipeline [22] | Implicit | Partial | Partial | Not addressed | Not energy-specific |
| Image acquisition pipeline [23] | Not addressed | Not addressed | Explicit for image data | Not addressed | No |
| CTA acquisition pipeline [24] | Not addressed | Partial | Explicit | Not addressed | No |
| SSN/FAIR/PROV concepts [9-10, 16] | Not addressed | Partial | Explicit | Partial | No direct case validation |
| Energy data protocols/DataPro [15, 19] | Partial | Partial | Partial | Partial | Yes, but less acquisition-specific |
| STREAM | Explicit | Explicit | Explicit | Explicit across four categories | Two industrial batch-process cases |

The classifications in Table 5 follow explicit coding rules. “Explicit” means that the reviewed approach contains a dedicated step, artifact, or rule for the criterion; “partial” means that the criterion is mentioned or supported by individual components but is not connected to an end-to-end acquisition decision; “implicit” means that the capability can be inferred but is not operationally specified; and “not addressed” means that no corresponding procedure was identified in the described scope. Industrial-energy validation indicates application scope, not superiority.

Prior approaches provide important building blocks, but STREAM combines them within one operational sequence. Compared with pipeline-oriented approaches, it adds an explicit objective and evidence gate before extraction. Compared with metadata and provenance standards, it adds indicator-specific requirements and suitability decisions. Compared with the energy data protocols and DataPro in [15, 19], it narrows the focus to acquisition readiness and adds stage-level failure actions, mandatory metadata, and a four-category uncertainty rubric

### 6.2 Contributions

The two cases show that the same six-stage procedure can be applied in different industrial data environments: IoT-based furnace data in the foundry case and combined SCADA and production-order data in the drying case.

A central finding is that data accessibility does not guarantee analytical suitability. In both cases, incomplete calibration information, heterogeneous sampling, weak synchronization, partial source-to-asset mapping, missing process context, and undocumented transformations affect dataset confidence.

Scientifically, STREAM contributes an operational integration of objective definition, requirement specification, source mapping, metadata augmentation, uncertainty assessment, and repository migration. Practically, it provides four outputs before indicators are used: a source map, an evidence record, a suitability classification, and a corrective-action list. These outputs help engineers distinguish limitations that can be managed through documentation or analytical restrictions from gaps that require system integration, configuration changes, or new measurements.

### 6.3 Implementation Implications

From an implementation perspective, STREAM should be applied before energy indicators are calculated or analytical models are developed. S and T define the decision, boundary, resolution, and tolerance conditions; R and E determine whether existing tags and records are directly suitable, conditionally usable, or require complementary evidence; and A and M preserve unknown values, quality flags, provenance, and processing status. In the two cases, this sequence supported immediate trend monitoring and exploratory analysis, while restricting high-confidence batch comparisons until calibration, synchronization, quantity, and contextual evidence could be strengthened.

The framework also supports staged data-infrastructure improvement. Some weaknesses can be resolved through documentation, while others require database or SCADA reconfiguration, sensor recalibration, or new measurements. Metadata such as source identifiers, tag names, units, data types, source paths, timestamp fields, and nominal sampling intervals may be extracted from machine-readable system descriptions. However, calibration status, installation context, process boundaries, and batch semantics generally require authoritative records or expert validation.

For prospective applications, implementation effort should be recorded by stage using person-hours, elapsed time, source systems inspected, candidate tags reviewed, and stakeholder consultations. Inter-rater consistency could be evaluated by asking independent practitioners to classify the same evidence package and reporting agreement statistics.

### 6.4 Limitations and Future Work

The evaluation provides quantitative diagnostic results but does not measure implementation effort, inter-rater agreement, changes in calculated indicators after STREAM-based alignment or exclusion, or the performance of automated metadata extraction. In addition, the cross-case implications remain mainly qualitative because comparable numerical records were unavailable for calibration completeness, temporal alignment, contextual linkage, and transformation provenance. STREAM identifies missing measurements, metadata, and contextual links but does not automatically resolve them, and the uncertainty assessment remains partly qualitative when numerical evidence is unavailable. Anonymized data and tag names also prevent exact numerical reproduction of the two cases. Nevertheless, the procedure can be replicated using the objective fields, technical requirements, source-map structure, uncertainty rubric, suitability rules, metadata template, and reporting format in Appendix Tables A1–A6.

Future work should prospectively record stage-level effort, assess inter-rater consistency, compare conventional and STREAM-compliant indicator calculations, and evaluate partial metadata extraction. It should also introduce quantitative readiness measures, including the proportion of variables meeting critical requirements, metadata completeness, timestamp-alignment error, contextual-linkage coverage, and the number of records affected by undocumented transformations. Further applications should cover additional sectors and continuous production processes.

## 7 Conclusion

This paper developed and evaluated STREAM, an objective-driven and uncertainty-aware framework for determining whether industrial sensor and production data are suitable for a specified energy-performance assessment. STREAM links the objective to indicators, required variables, physical and digital sources, metadata, uncertainty records, and analytical repositories through six stages.

The framework is evaluated through induction-furnace melting and cheese-powder drying. In both cases, STREAM supports variable identification, source mapping, metadata documentation, uncertainty assessment, and data-warehouse migration. The results show that accessible data can remain unsuitable for reliable batch-level analysis when calibration evidence, timestamp alignment, batch linkage, process context, or transformation history is incomplete.

The scientific contribution is an integrated acquisition framework combining objective-driven requirements, source-suitability assessment, metadata archival, repository migration, and uncertainty documentation. The practical contribution is a repeatable procedure for diagnosing data readiness, identifying measurement and integration gaps, and preparing traceable datasets for monitoring, benchmarking, and decision support.

The study is limited by quantitative uncertainty assessment, reliance on domain expertise, and two anonymized batch-process cases. Future work should quantify uncertainty propagation, evaluate implementation effort, test the framework across additional sectors and continuous processes, and develop software support for automated metadata extraction, real-time data-quality assessment, and uncertainty-aware analytics.

**Acknowledgments.** This work is part of the project "Data-driven best-practice for energy-efficient operation of industrial processes - A system integration approach to reduce the CO2 emissions of industrial processes" (project no.64020-2108) by the Energy Technology Development and Demonstration (EUDP) program, Denmark; Part of the project titled "Danish Participation in IEA IETS Task XVIII Digitalization, Artificial Intelligence and Related Technologies for Energy Efficiency and GHG Emissions Reduction in Industry Subtask 4", funded by EUDP (project number: 34251-549157); Part of the project titled "Danish participation in IEA IETS Task XXII - Climate Resilience and Energy Adaptation in Industry under Uncertainty", funded by EUDP (project number: 134243-534852).

**Disclosure of Interests.** The authors have no competing interests to declare that are relevant to the content of this article.

## Appendix

The Appendix provides detailed metadata templates, uncertainty rubrics, and case-specific traceability matrices that support the compact presentation in the main text.

**Table A1.** Metadata structure required for traceable and uncertainty-aware energy-data acquisition.

| Metadata category | Included attributes |
|---|---|
| Dataset metadata | Dataset identifier, description, owner, creation time, retention period, access rights, and version |
| Sensor metadata | Sensor identifier, sensor type, manufacturer, measurement range, accuracy, resolution, device version, and installation date |
| Asset metadata | Asset identifier, equipment type, process area, installation location, and sensor-to-asset relationship |
| Measurement metadata | Measured quantity, unit, expected range, expected sampling interval, reference time zone, and boundary definition |
| Calibration metadata | Calibration date, procedure, validity status, and available uncertainty estimate |
| Process-context metadata | Batch identifier, recipe or product family, material quantity, operating state, and relevant start/end events |
| Quality metadata | Validity status, missing-data status, duplicate status, anomaly flag, sampling status, and clock-synchronization status |
| Provenance metadata | Source system, extraction method, extraction time, transformation history, software or script version, and responsible actor |
| Uncertainty metadata | Uncertainty category, source of uncertainty, numerical estimate or qualitative level, unit, confidence level, and mitigation action |

**Table A2.** Data-quality risk assessment criteria.

| Category | Evidence assessed | Low risk | Moderate risk | High or unknown risk |
|---|---|---|---|---|
| Measurement | Calibration, accuracy, resolution, drift, and installation context | Calibration is current, and sensor accuracy meets the analytical requirements. | Minor evidence gaps or deviations exist but do not compromise the intended analysis. | Critical calibration or accuracy evidence is missing, expired, inconsistent with the installation context, or unknown. |
| Temporal | Sampling interval, clock source, timestamp alignment, and batch boundaries | Critical timestamps are synchronized within the required tolerance. | Minor timing deviations can be documented, corrected, or accounted for analytically. | Sampling or clock mismatches affect allocation to batches or orders, or the timestamp source is unknown. |
| Contextual | Asset, order, recipe, material, equipment state, and process boundary | Observations are linked to the relevant asset, batch or order, product, and operating state. | Partial linkage is available, with assumptions explicitly documented. | Critical contextual information is missing, ambiguous, unverifiable, or inconsistent. |
| Processing | Filtering, resampling, interpolation, aggregation, and unit conversion | Raw data are preserved, and all transformations are fully documented. | Minor transformations are documented and appropriate for the intended use. | Transformations are undocumented, irreversible, potentially bias derived indicators, or their history is unknown. |

**Table A3.** Variable-source traceability and suitability assessment for Case Study 1.

| Variable group | Source and evidence required | Uncertainty and suitability | Recommended action |
|---|---|---|---|
| Electricity consumption | IoT energy meter or control-system energy record, with meter ID, unit, calibration status, sampling interval, and source timestamp. | Conditional suitability due to measurement and temporal uncertainty. | Verify calibration and align energy intervals with batch boundaries. |
| Furnace temperature | Furnace temperature sensor, with sensor ID, location, unit, calibration status, and sampling interval. | Conditional suitability due to measurement and contextual uncertainty. | Verify sensor location and calibration status. |
| Material quantity | Charge or material records, with batch ID, mass unit, weighing method, and material category. | Conditional to limited suitability due to contextual and measurement uncertainty. | Strengthen batch-material linkage and verify mass evidence. |
| Processing time and batch boundaries | IoT timestamps and control-state events, with start and end definitions, clock source, and furnace ID. | Conditional suitability due to temporal and contextual uncertainty. | Synchronize timestamps and document batch-boundary rules. |
| Equipment status | Control tags or inferred state records, with state definitions, tag-to-asset mapping, and transformation logic. | Conditional suitability due to contextual and processing uncertainty. | Validate state logic and preserve raw state-change records. |

**Table A4.** Data-uncertainty assessment in Case Study 1.

| Uncertainty category | Case example | Effect | STREAM treatment |
|---|---|---|---|
| Measurement | Incomplete calibration information for electricity and temperature sensors | Reduced confidence in measured and derived values | Record accuracy, calibration status, and validation evidence in metadata |
| Temporal | Different sampling intervals and uncertain batch boundaries | Potential misallocation of energy to melting batches | Assess timestamp alignment and document boundary rules and tolerance |
| Contextual | Incomplete links among sensor records, furnace identity, batch records, and operating state | Uncertain batch segmentation and limited comparability | Link observations to furnace, batch, and process-state metadata |
| Processing | Resampling, interpolation, aggregation, or unit conversion | Possible distortion of derived indicators | Preserve raw data and document all transformation methods and assumptions |

**Table A5.** Variable-source traceability and suitability assessment for Case Study 2.

| Variable group | Source and evidence required | Uncertainty and suitability | Recommended action |
|---|---|---|---|
| Electricity demand | SCADA power or energy tags, with tag ID, unit, sampling interval, meter or calculation basis, and timestamp. | Conditional suitability due to measurement and temporal uncertainty. | Verify the metering basis and align demand data with order periods. |
| Heating demand | Heating records or SCADA-derived heating data, with measurement method, unit conversion, and validation evidence. | Conditional to limited suitability due to measurement and processing uncertainty. | Improve heating-data quality and document the derivation procedure. |
| Temperature, humidity, and pressure | SCADA process sensors, with sensor location, unit, calibration status, sam- | Conditional suitability due to measurement and contextual uncertainty. | Verify sensor context, calibration sta- |
| | pling interval, and process zone. | | tus, and sampling consistency. |
| Feed and product quantity | Production or order database, with order ID, product mass, feed rate, recipe, and product family. | Conditional suitability due to contextual and measurement uncertainty. | Strengthen linkage among quantity, recipe, and order records. |
| Batch timing | Order database and SCADA timestamps, clock source, and time zone. | Conditional suitability due to temporal and contextual uncertainty. | Synchronize SCADA and order records and document timing rules. |
| Equipment status and data transformations | SCADA state tags and processing scripts, with state definitions, raw values, and resampling or aggregation methods. | Conditional suitability due to processing and contextual uncertainty. | Preserve raw data, validate state definitions, and document all transformations. |

**Table A6.** Data-uncertainty assessment in Case Study 2.

| Uncertainty category | Case example | Effect | STREAM treatment |
|---|---|---|---|
| Measurement | Incomplete calibration information for energy and operational sensors; low quality of heating data | Reduced confidence in measured and derived energy values | Record accuracy, calibration status, validation evidence, and heating-data limitations |
| Temporal | Different sampling rates across SCADA and order data | Misalignment of energy use and production batches | Assess timestamps and document synchronization and boundary uncertainty |
| Contextual | Incomplete links to orders, recipes, product quantities, and operating states | Limited comparability across drying batches | Link observations to orders, recipes, assets, quantities, and process states |
| Processing | Resampling, interpolation, aggregation, or unit conversion | Potential bias in derived indicators | Preserve raw data and document all transformations and assumptions |